\documentclass{article}

\usepackage[english]{babel}
\usepackage{upgreek}
\usepackage{amsmath, amssymb}

\usepackage[letterpaper,top=2cm,bottom=2cm,left=3cm,right=3cm,marginparwidth=1.75cm]{geometry}

\usepackage{amsmath}
\usepackage{graphicx}
\usepackage[colorlinks=true, allcolors=blue]{hyperref}

\title{Your Paper}\title{Nonreciprocal multi-body interactions activate liquid state of acoustically levitated particle ensembles}
\author{Nina M.\ Brown, Heinrich M. Jaeger}
\begin{document}

\maketitle

\begin{abstract}
Nonreciprocal forces are often a consequence of asymmetry in the properties of the interacting objects.  
However, even if all objects are identical and isotropic, and the pairwise interactions between two objects are completely reciprocal, nonreciprocal forces can still appear when an arrangement of many objects breaks configurational symmetry in the presence of non-pairwise, multi-body interactions.
Here we demonstrate that such emergent nonreciprocity can activate a particle ensemble to behave like a liquid, albeit with unique traits.  In our experiments, passive microspheres are acoustically levitated in air, where they form a freely floating monolayer containing up to a couple hundred particles and collectively behave like a two-dimensional liquid droplet. The particles interact via nonreciprocal multi-body forces that arise from the combination of acoustic scattering and sound-induced viscous microstreaming. We find that these forces drive superdiffusive particle motion with non-Gaussian tails in the particles’ speed distribution.  Using probes that reach laterally into the levitation plane, we perform liquid pendant drop and pinch-off experiments. Compared to ordinary liquids, the droplets are found to have a kinematic viscosity similar to that of water, but in combination with an extremely low interfacial tension.  The pinching-off is driven by nonreciprocity-induced active fluctuations and exhibits the self-similar double-cone neck profile seen also in liquid nanojets close to rupture, however here characterized by power law behavior with a scaling exponent that is anomalously small.  
\end{abstract}

\newcommand{\p}[1]{\left ( #1 \right )}
\newcommand{\br}[1]{\left [ #1 \right ]}
\newcommand{\ol}[1]{\overline{ #1 }}
\newcommand{\Bo}{\text{Bo}~}
\newcommand{\mb}[1]{\mathbf{ #1 }}
\newcommand{\red}[1]{\textcolor{red}{#1} }
\newcommand{\blue}[1]{\textcolor{blue} {#1}}

\maketitle

\section{Introduction}
In systems ranging from active matter to robotic metamaterials to particle-laden plasmas, nonreciprocal forces are known to induce dynamics and spatio-temporal patterns not observed with conservative interactions \cite{Fruchart_Vitelli_2026, Dinelli_OByrne_Curatolo_Zhao_Sollich_Tailleur_2023, Fruchart_Hanai_Littlewood_Vitelli_2021, Marchetti_2013}. 
Nonreciprocal forces typically arise from an explicit symmetry breaking, either because of an asymmetry in the properties of the interacting species or objects \cite{King_Morrell_Sustiel_Gronert_Pastor_Grier_2025},
 or because the nonreciprocity has been programmed into the interactions, as in robotic swarms or metamaterials \cite{Fruchart_Hanai_Littlewood_Vitelli_2021}. 
Nonreciprocal interactions can also occur among a single species or identical objects if there is a front-back asymmetry, such as a limited vision cone or a trailing wake \cite{Matthews_Sanford_Kostadinova_Ashrafi_Guay_Hyde_2020, Yu_Abdelaleem_Nemenman_Burton_2025, Shu_Li_Yu_Burton_2026}. 
A rather different situation occurs in systems comprising identical objects with pairwise forces that obey reciprocity.  In this case, nonreciprocal forces require non-pairwise, i.e., multi-body interactions.  This has been studied, in particular, for multi-body interactions mediated by a field. 
Examples include active particles in a temperature or chemical concentration field \cite{Zampetaki_Liebchen_Ivlev_Lowen_2021}, and passive particles in an optical \cite{Davenport_Kleckner_2022, Li_Liu_Lin_Ng_Chan_2021, Parker_Nagasamudram_Peterson_Li_Soleimanikahnoj_Rice_Scherer_2025}
or a sound \cite{StClair_Davenport_Kim_Kleckner_2023, Morrell_Elliott_Grier_2026, Wu_VanSaders_Lim_Jaeger_2023, Wu_2025_3body} field.  
When asymmetries in the spatial configuration of the particles distort the field, this then can lead to net momentum flux and generate movement.  Such emergent nonreciprocity tends to amplify spatial particle fluctuations via positive feedback. 
An intriguing outcome is that  even if all particles individually are  passive and non-motile, activity can be induced by nonreciprocal multi-body interactions.  
In small groups of identical passive, isotropic particles, this induced activity has been found to enable collective propulsion and also limit cycles \cite{Davenport_Kleckner_2022, Li_Liu_Lin_Ng_Chan_2021, StClair_Davenport_Kim_Kleckner_2023, Wu_2025_3body}. 
In larger ensembles, it has been observed to drive intermittent local rearrangements that, with increasing level of activity, destroy ordered particle configurations and eventually give way to a liquid-like state \cite{ Wu_VanSaders_Lim_Jaeger_2023}.  
However, the macroscale properties of such a nonreciprocal liquid have not yet been studied.

Here we report on experiments that enable us to extract properties such as an effective line tension and viscosity from two-dimensional (2D) particle liquids activated by nonreciprocal multi-body interactions. 
We use acoustic levitation in air to lift microparticles against gravity and self-assemble them into monolayer configurations in a well-defined levitation plane. 
Within that plane, the particles interact via forces generated by an intense ultrasound field, forming 2D ensembles that behave like liquid droplets.
The experiments are enabled by a unique setup in which levitating particle ensembles can be manipulated with probes that reach into an acoustic cavity \cite{Brown_VanSaders_Kronenfeld_DeSimone_Jaeger_2024_RSI,Brown_VanSaders_Kronenfeld_DeSimone_Jaeger_2024_GM}. 
Suitably designed probes make it possible not only to generate uniform  acoustic potential gradients within the levitation plane, but also to apply controlled tensile stress to the droplets.
During the experiments, the motion of all particles can be tracked with high-speed video.

\begin{figure}[h]
\centering
\includegraphics[width=\linewidth]{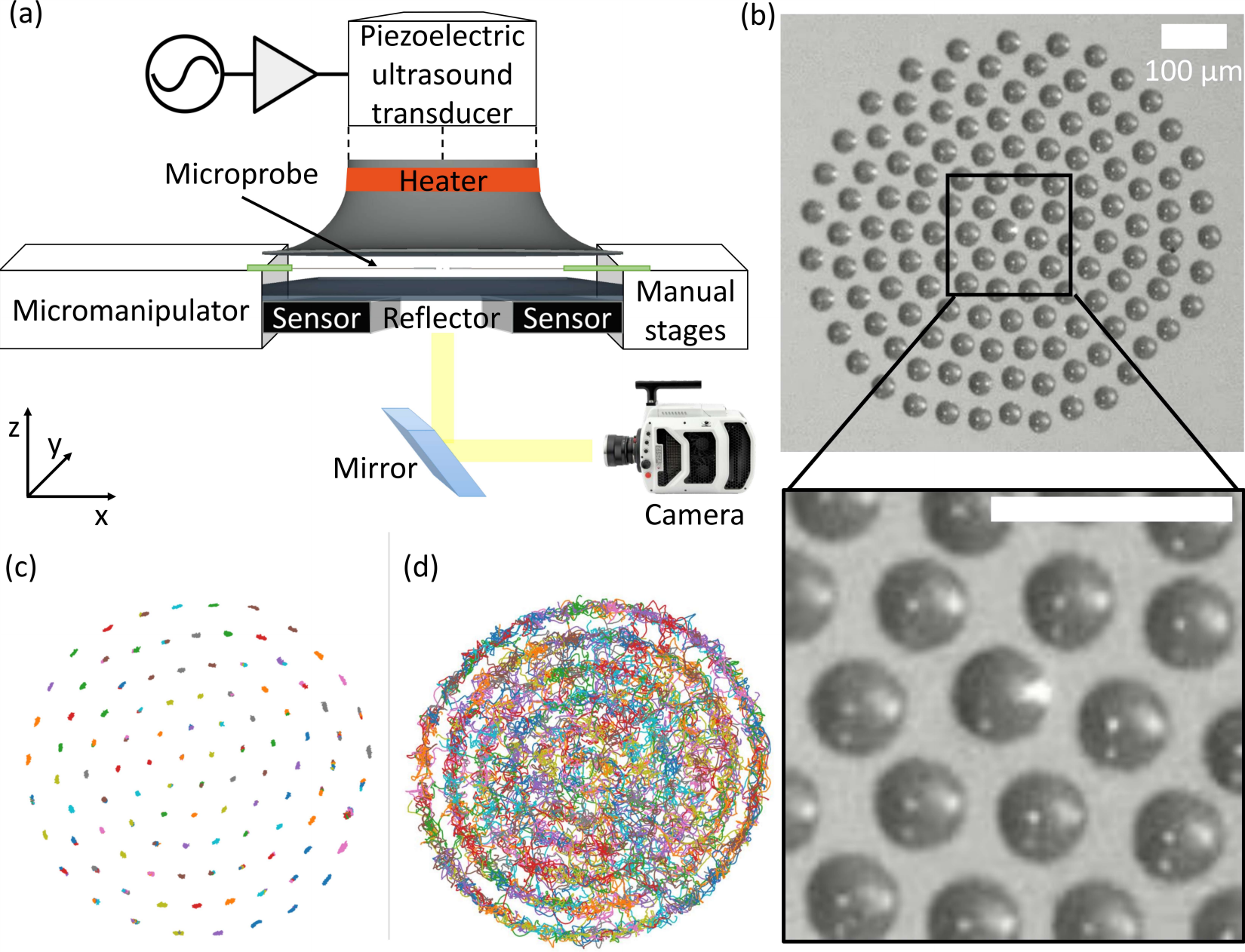}
\caption{
Experimental realization of an acoustically levitated, 2D liquid of identical granular particles.
(a) Sketch of the setup, including micromanipulators with probes that reach into the acoustic cavity to perturb levitating particle liquid droplets.
(b) Snapshot from high-speed video of a droplet of microparticles (top), with zoomed-in central section (bottom).
Scale bar: 100 $\upmu$m.
(c) Particle trajectories over 0.1 s within a droplet at a relatively low sound pressure of $p_{\mathrm{ac}} = 3880$ Pa.
(d) Same as (c) but for $p_{\mathrm{ac}} = 5920$ Pa.
}
\label{fig:acousticlevitation}
\end{figure}

Figure \ref{fig:acousticlevitation}(a) shows a sketch of our experimental setup. A piezoelectric ultrasound transducer (40.45 kHz) drives an aluminum Langevin horn positioned half a sound wavelength ($\lambda= 8.5$ mm) above a reflecting glass surface, causing a standing wave to form.
This creates an acoustic resonance cavity with a low-pressure zone at the central nodal plane where particles can be levitated against gravity.
Our experiments use monodisperse polystyrene spheres with diameter $d=$ 40.4 $\upmu$m << $\lambda$, i.e., we operate deep in the Rayleigh scattering limit.
We control the acoustic energy density \cite{Lim_VanSaders_Jaeger_2024} 
\begin{equation}
    E_0 
    = 
    \frac{1}{2} \beta_0 p_{\mathrm{ac}}^2
    \label{eq:E0}
\end{equation}
in the cavity by monitoring it with ultrasound pressure sensors, adjusting the amplitude of the signal applied to the ultrasound transducer, and by stabilizing the temperature of the horn \cite{Brown_VanSaders_Kronenfeld_DeSimone_Jaeger_2024_RSI}.
Here, $\beta_0$ is the compressibility of the levitation medium (air), and $p_{\mathrm{ac}}$ is the RMS amplitude of the oscillating sound pressure.
Experimental values of $p_\text{ac}$ are estimates of the maximum pressure in the levitation plane determined from measurements of the system combined with finite element analysis simulations (see SI).
The levitating particle rafts are imaged from below with a high-speed camera at rates of 5,000-10,000 frames per second (Fig.\ \ref{fig:acousticlevitation}(b)).

The particle separation within the levitating droplets, which results from the balance between attractive forces from scattered sound and repulsive forces from sound-induced microstreaming, is controlled by the Stokes number $\Omega=\omega d^2/(4\nu) = (d/\delta_{\nu})^2/2$  \cite{Lim_VanSaders_Jaeger_2024, Wu_VanSaders_Lim_Jaeger_2023}. Here $\omega$ is the sound frequency, $d$ is the particle diameter, $\nu$ is the kinematic viscosity of the levitating medium (air), and $\delta_{\nu} = \sqrt{2\nu/\omega}$ is the characteristic thickness of the viscous boundary layer surrounding the particles, which effectively generates an enlarged, hydrodynamic particle radius.
For our particle size and sound frequency, $\delta_{\nu} \approx 11~\upmu$m.
This gives $\Omega \simeq 6.6$ and leads to a center-to-center spacing of 1.26$d$ between particles in the levitation plane.
Since the competing forces are both proportional to the sound intensity given by Eq.\ \ref{eq:E0}, this particle spacing is independent of $E_0$ or, equivalently,  $p_{\mathrm{ac}}$ \cite{Lim_VanSaders_Jaeger_2024}.

\section{Acoustically-driven active liquid state}
As the the sound pressure $p_{\mathrm{ac}}$ is increased beyond a certain level,  in-plane particle fluctuations due to non-conservative forces drive a levitating particle ensemble from a nearly quiescent, highly ordered state to a state of constant particle movement, resembling a liquid droplet \cite{Wu_VanSaders_Lim_Jaeger_2023}. Figure \ref{fig:acousticlevitation}(b) shows a snapshot of such a droplet, while (c, d) show individual particle trajectories for a comparable particle raft over 0.1 s at different $p_\text{ac}$.
Video of the experiments portrayed in (c, d) can be seen in Movie S1.
As the liquid-like state is entered, structural ordering first diminishes in the outer layers of the droplet, where there are more local packing defects relative to the droplet core due to the droplet's round boundary.

\begin{figure}[h]
\centering
\includegraphics[width=\linewidth]{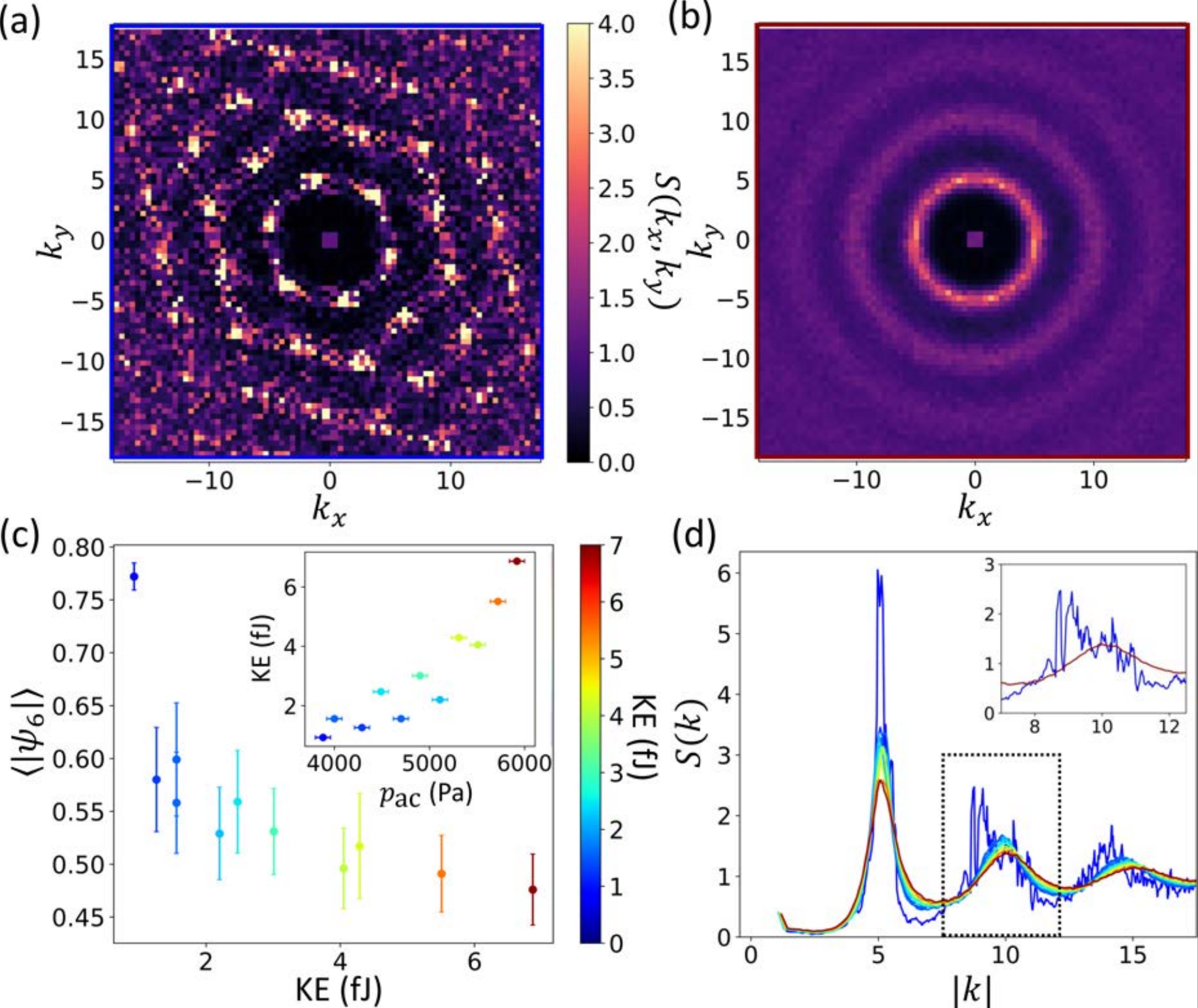}
\caption{ Structural characterization of 2D liquid droplets.
    (a) Static structure factor $S(k_x, k_y)$ at the lowest sound pressure probed (3880 Pa)).
    (b) Same as (a) but at the highest pressure (5920 Pa).
    (c) Orientational order parameter $|\psi_6|$  and average kinetic energy $KE$ of particles within droplets as a function of acoustic pressure $p_{\mathrm{ac}}$ (error bars: standard deviation over time of mean $|\psi_6|$).
    (d) Static structure factor magnitude $S(k)$ vs $|k|$ for different pressures $p_{\mathrm{ac}}$.
    (d, inset) $S(k)$ vs $|k|$ for the highest and lowest acoustic pressures, zoomed to the second peak. 
    }
\label{fig:thermal}
\end{figure}

This change in behavior is reflected in the static structure factor $S(k)$, shown in Fig.\ \ref{fig:thermal}(a, b).
For a solidlike droplet at low $p_{\mathrm{ac}}$, $S(k_x, k_y)$  displays the hexagonal structure associated with a close-packed particle configuration, while for higher pressure this pattern turns into circular rings, indicating a complete lack of orientational order.
The loss of orientational order with increasing $p_{\mathrm{ac}}$ can be quantified by the orientational order parameter $|\psi_6|$ (see methods), which is plotted in Fig.\ \ref{fig:thermal}(c).
As $|\psi_6|$ decreases, the same plot shows how the average kinetic energy per particle increases rapidly with $p_{\mathrm{ac}}$. 
Kinetic energy data are plotted both for increasing and decreasing $p_\text{ac}$.
The variation in the data is due to a slight change in the experimental environment over time; closed-loop feedback control of the acoustic pressure helps to mitigate this, but it is difficult to eliminate entirely.
In the following, we take this kinetic energy as a proxy for a particle-motion-based effective temperature of the droplets. 

In 3D systems, the Hansen-Verlet rule classifies a material as transitioning from a solid-like to a liquid state once the first peak of $S_\text{max}(k) > 2.85$ \cite{Hansen_Verlet_1969}. 
In 2D, the equivalent criterion is $S_\text{max}(k) \gtrsim 4-5$ \cite{Wang_Alsayed_Yodh_Han_2010}, where different interparticle interactions result in slightly different critical values, e.g.\  4.4 for a plasma \cite{Caillol_Levesque_Weis_Hansen_1982} and 4.7 for a Lennard-Jones fluid \cite{Ranganathan_1992}.
In our data, the first peak of $S(k)$ decreases from 6.0 very quickly to 3.5 and then to 2.6 with increasing $p_{\mathrm{ac}}$ (Fig.\ \ref{fig:thermal}(d)). 
This suggests that all droplets tested should be in a liquid state, except for the droplet at the lowest $p_\text{ac}$ value.
However, the height of the first peak of $S(k)$ in systems with small number of particles $N$ is known to be lower than for large $N$ \cite{Zhuravlyov_Goree_Elvati_Violi_2023}. 
This means that our rafts, with $N \sim 100$, enter their liquid state at the low end of the pressure range shown in Fig.\ \ref{fig:thermal}(d).
This is supported by the fact that at $p_{\mathrm{ac}} =$ 3880 Pa, where the first peak of $S(k) = 6.05$, the hexagonal pattern in $S(k_x, k_y)$ still is dominant.
Furthermore, at the lowest $p_{\mathrm{ac}}$, the second peak of $S(k)$ exhibits two major subpeaks (Fig.\ \ref{fig:thermal}(d, inset)), while curves for higher $p_{\text{ac}}$ do not.
Splitting of the second peak is expected once a freezing transition is crossed  \cite{Wang_Alsayed_Yodh_Han_2010}.
We find closely related trends with acoustic pressure in the radial distribution function $g(r)$ (see SI).

\begin{figure}[h]
\centering
\includegraphics[width=\linewidth]{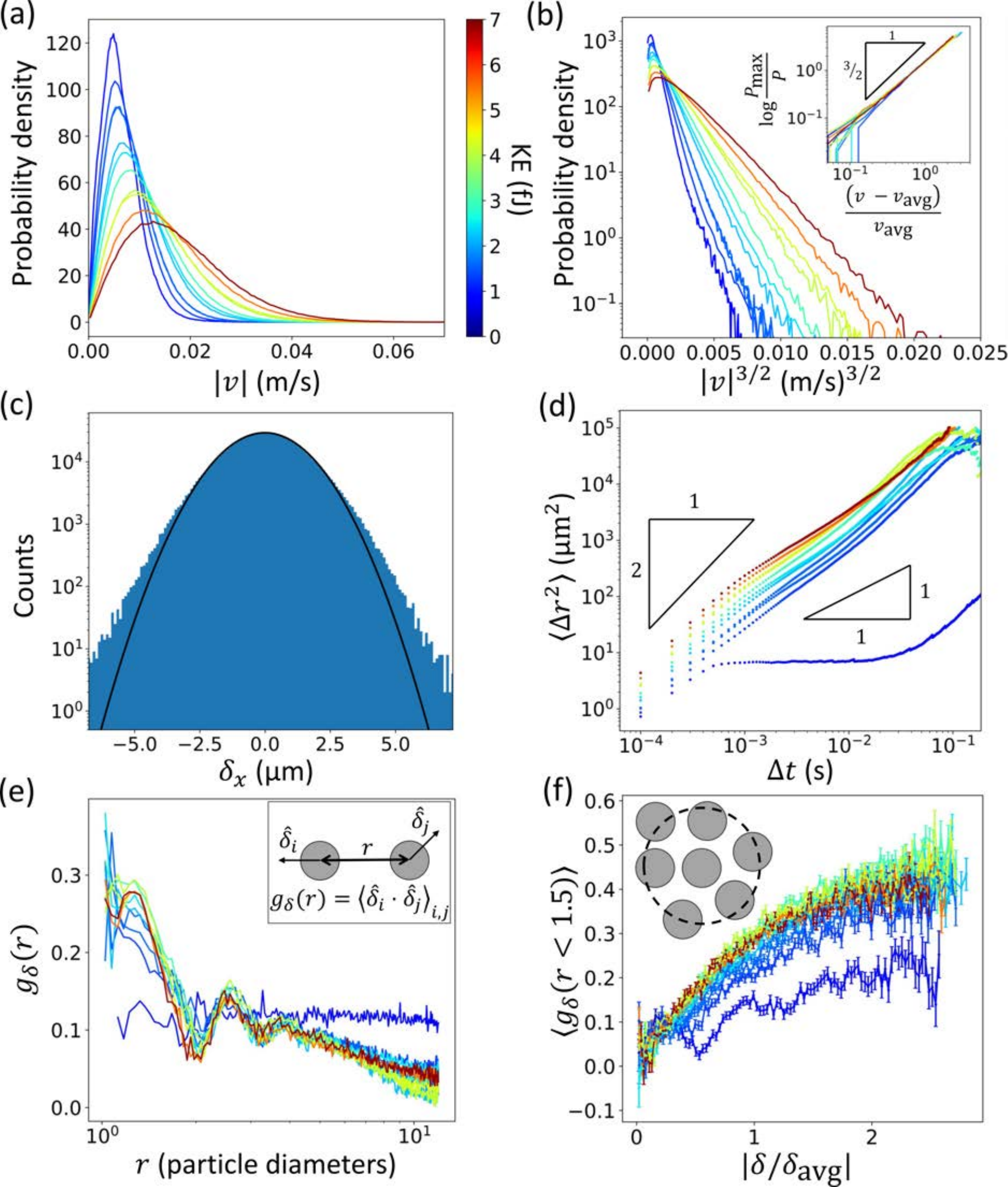}
\caption{
    Statistics of particle motion for a levitated granular droplet ($N = 106$) at different $p_\text{ac}$.
    (a) Speed distributions for the droplet, colored by mean particle kinetic energy.
    Colorbar also applies to (b, d-f).
    (b) Speed distributions vs.\ $|v|^{3/2}$, $y$-axis on a log scale. 
    (b, inset) Normalized log of the distribution tails vs.\ $(v - v_{\text{avg}})/v_{\text{avg}}$, on a log-log scale. 
    Triangle drawn with slope $3/2$.
    (c) Distribution of particle displacements in one direction for a high-pressure experiment ($p_\text{ac} = 5920$ Pa), $y$-axis on a log scale. 
    Line is a Gaussian fit.
    (d) Mean square displacement vs.\ time on a log-log scale.
    Triangles are drawn with slopes 1 and 2.
    (e) Displacement correlation function $g_\delta(r)$ vs.\ $r$, $x$-axis on a log scale. 
    (e, inset) Schematic of $g_\delta(r)$ for two particles.
    (f) Mean displacement correlation function $g_\delta$ value within a small neighborhood of each particle vs.\ normalized displacement distance (error bars: standard error).
    (f, inset) Schematic of particles with a dashed circle drawn at a distance of $1.5a$ from the central particle. 
    $\langle g_\delta (r < 1.5) \rangle$ for the central particle is the mean $g_\delta(r)$ computed between the central particle and the particles with their centers within the circle.
    }
\label{fig:disvel}
\end{figure}

\textit{Speed distributions and anomalous diffusion}. Figure\ \ref{fig:disvel} shows results from tracking individual particle motion inside  a droplet of $N = 106$ particles at different $p_\text{ac}$.
The parameters measured in Fig.\ 3 are common to all $N$ probed.
The speed (a, b) and displacement (c) distributions differ from the thermal case by exhibiting wider tails.
To highlight this, in Fig.\ \ref{fig:disvel}(b) we plot  the speed distributions against $|v|^{3/2}$.
An exponential distribution $P \sim e^{- x^k}$ with exponent $k = 3/2$ is seen to provide a remarkably good fit not only to the high-speed tails but also at lower speeds closer to the peak, i.e., the average kinetic energy $KE$.
Still, an even better match is achieved with a slightly smaller exponent.
In the inset of Fig.\ \ref{fig:disvel}(b), the same data are plotted as the scaled  probability  $\log \br{P_\text{max}/P}$ versus $(v - v_\text{avg})/v_{\text{avg}}$ for $v > v_\text{avg}$.
The slopes on this plot represent the exponent $k$.
From a best fit to each curve, we obtain $k = 1.34 \pm 0.06$ (mean $\pm$ SD).
This lies in contrast to a thermally activated system, which follows a Gaussian distribution ($k=2$).

Exponents $k < 2$ are seen in some other granular systems.
For a thermally heated granular gas with collisions, it is predicted that the tail of the velocity distribution goes like $k=3/2$ \cite{van_Noije_Ernst_1998}.
However, our distributions differ from a Gaussian over most of their width, not just the tails.
Experiments of some granular systems have also demonstrated distributions that  have $k = 3/2$ or other deviations from a Gaussian that persist across the full range of velocities rather than just in the tail \cite{Rouyer_Menon_2000, Prasad_Das_Sabhapandit_Rajesh_2019}.
Alternate theories have tried to reconcile this, incorporating features such as randomly varying coefficients of restitution \cite{Barrat_Trizac_2003} and negative friction \cite{Demaerel_De_Roeck_Maes_2020}.
In our experiments, the hard-sphere particles are surrounded by a soft shell formed by a repulsive viscous boundary layer set up by microstreaming flows \cite{Lim_VanSaders_Jaeger_2024, Wu_2025_3body}.  
Random values for the coefficient of restitution might then be a consequence of particle collisions at different impact parameters, or perhaps the nonreciprocal nature of the interactions results in an effective negative friction.

Clear deviations from the Gaussian behavior expected for thermal  motion are also seen in the particle displacement distributions. Figure \ref{fig:disvel}(c) shows a representative histogram of displacements parallel to the $x$-axis between sequential imaging frames ($\Delta t = 1/10000$ s).
Thus, particles within an acoustically levitated liquid droplet are more likely to move at high velocities (large displacements) relative to molecules comprising a non-interacting thermal fluid.
This translates into superdiffusive behavior, whereby the mean square displacement (MSD) increases with time faster than linear (Fig.\ \ref{fig:disvel}(d)). 
This superdiffusion is typical for systems of active particles \cite{Chen_Wei_Sheng_Tsao_2016, Ribeiro_Potiguar_2016}, living organisms \cite{Ariel_Rabani_Benisty_Partridge_Harshey_Be’er_2015}, and other systems with nonreciprocal forces like particle-laden (`dusty') plasmas \cite{Juan_I_1998, Andrew_Guazzotto_Matthews_Hyde_Kostadinova_2025}.
The MSD of the solid-like droplet probed displays an initial steeper slope, which quickly falls off to a flat line; this suggests that the particles experience small fluctuations on a short timescale, but are caged in a lattice configuration.
Once in a liquid state, even at the lowest effective temperatures, particles  do not appear to experience significant caging.
This is consistent with the lack of caging observed for particle rafts in an intermediate state that displays intermittent dynamics \cite{Wu_VanSaders_Lim_Jaeger_2023}.

To further check for vestiges of the string-like collective particle displacements observed as intermittent disruptions of the solid-like state at lower sound pressure \cite{Wu_VanSaders_Lim_Jaeger_2023}, we define a  correlation parameter for simultaneous displacement between two particles as 
\begin{equation}\label{eq:gdelta}
    g_\delta(r)
    =
    \langle \hat{\delta}_i \cdot \hat{\delta}_j \rangle_{i,j}.
\end{equation}
Here $\hat{\delta}_i$ is the unit vector in the direction of the displacement of particle $i$, and the average is taken over all particles in a droplet.
A schematic is drawn in the inset of Fig.\ \ref{fig:disvel}(e).
This quantifies the degree to which a particle's direction of motion is correlated with that of  particles a distance $r$ away.
Fig.\ \ref{fig:disvel}(e) shows  $g_\delta(r)$ averaged over the course of each experiment.
The dark blue curve that remains relatively flat represents the solid-like droplet. 
In the solid case, the correlation is small but positive, suggesting limited relative particle motion with some long-time global motion (rotation) of the entire raft.
Each of the remaining curves for the liquid-like droplets exhibits peaks at the first two neighbor positions before gradually decreasing with a trend that appears logarithmic.
Even when particles are closer than their mean spacing ($r < 1.26 d$), $g_\delta$ tends to be positive.
Within these droplets, one particle that moves close to another may cause the other particle to move away, but the first particle does not necessarily move in the opposite direction; this may be a result of the high velocity required for particles to be in closer proximity than the distance imposed by their viscous boundary layers.
We also performed molecular dynamics simulations of a 2D Yukawa crystal with Brownian fluctuations and saw very little correlation of particle displacements beyond close-range negative correlations for strong interparticle interactions (see SI).

Figure \ref{fig:disvel}(f) demonstrates collective motion of neighboring particles.
The plot shows the mean $g_\delta$ value within $1.5d$ of a particle $\langle g_\delta(r<1.5) \rangle$ plotted against the normalized displacement magnitude $\left | \delta / \delta_\text{avg} \right |$.
The inset is a diagram of a particle and its small neighborhood; $g_\delta$ is calculated between the central particle and each other particle in the circle then averaged.
For all $KE$, $\langle g_\delta(r<1.5) \rangle$ is small for small displacements and increases for larger displacements.
This effect is less significant for the solid-like case.
When particles are moving quickly, they are likely to be moving in the same direction as their neighbors.
Effectively, this collective motion is akin to flocking of birds \cite{ Cavagna_Cimarelli_Giardina_Parisi_Santagati_Stefanini_Viale_2010, Kumar_Soni_Ramaswamy_Sood_2014} and is a signature of active and nonreciprocal systems \cite{Fruchart_Hanai_Littlewood_Vitelli_2021}.

Experiments discussed in the remainder of this manuscript were performed with droplets in a liquid-like state.

\section{Pendant droplets and interfacial tension}

\begin{figure}[h]
\centering
\includegraphics[width=\linewidth]{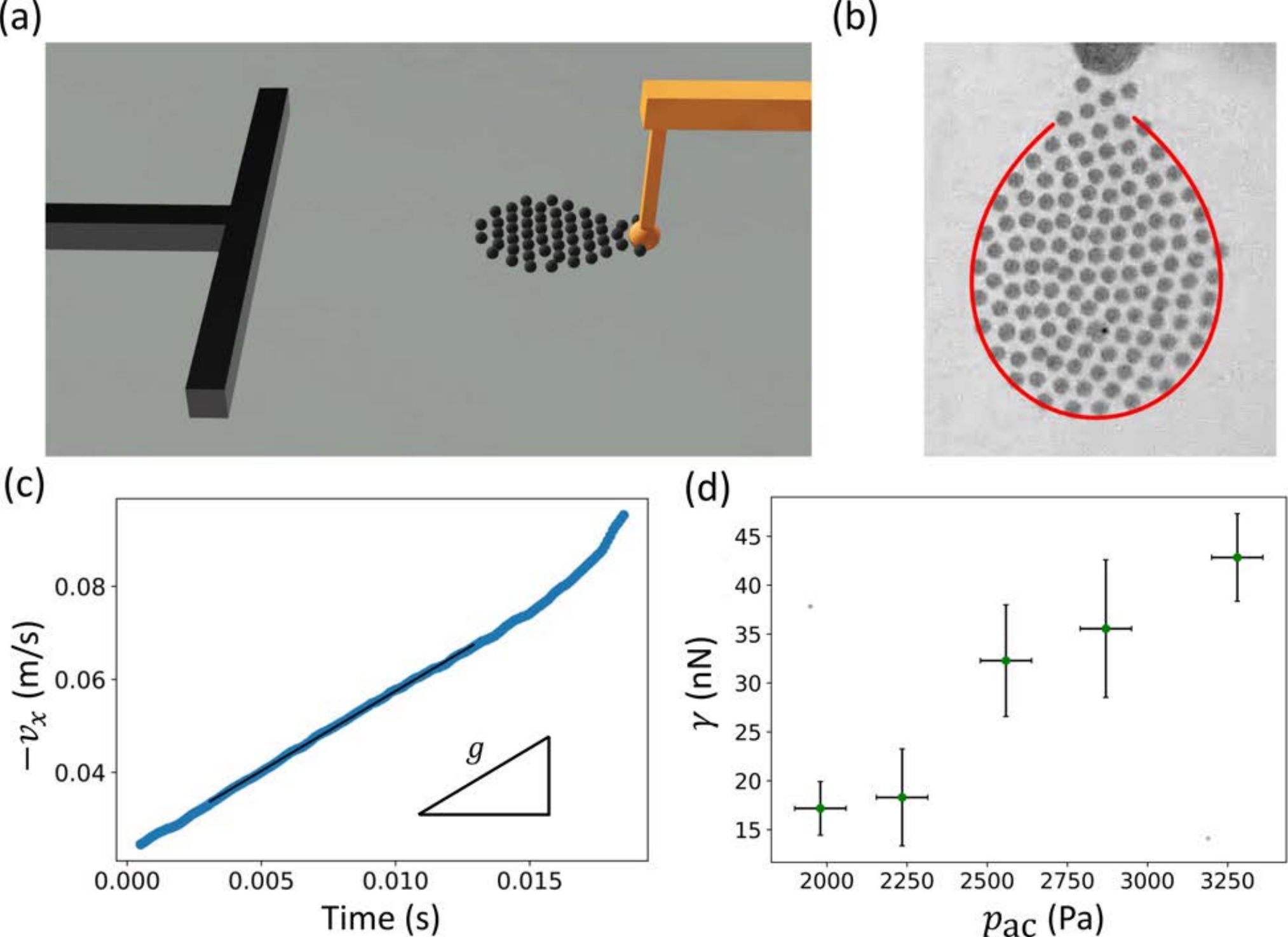}
\caption{
    Pendant drop experiments were performed to quantify the interfacial tension of this granular liquid. 
    (a) Experiment diagram.
    (b) Experiment image with drop profile drawn (red).
    (c) Mean particle velocity while droplet is ``falling'' between the two probes. A line is fit to a segment of the data; triangle shows slope (droplet acceleration).
    (d) Surface tension vs.\ effective temperature parameter $p_\text{ac}$.
    }
\label{fig:pendantdrop}
\end{figure}

Pendant drop experiments are commonly used to extract the surface tension $\gamma$ of liquids.
In such experiments, a small droplet of the liquid  hangs from a surface or needle and deforms under the influence of gravity, which makes it possible to extract $\gamma$ from analyzing the droplet profile.
The droplet shape is determined by the balance of surface tension $\gamma$ and gravity, which is expressed by the Bond number $\Bo$defined as
\begin{equation}\label{eq:Bondgrav}
    \Bo
    =
    \frac{\Delta \rho g R_0^2}{\gamma}.
\end{equation}
Here, $\Delta \rho$ is the difference in density between the droplet  and the external medium (often air), $g$ is gravitational acceleration, and $R_0$ is the radius of the droplet's bulb.
The expression for the Bond number turns out to be the same for the 2D case, but now $\Delta \rho$ refers to area density and $\gamma$ becomes a line tension (see SI).
By subjecting a levitating droplet to a uniform force field that mimics gravity, we can therefore extract the Bond number from measurements of the droplet shape, which provides access to the line tension.

A side view schematic of our experimental setup is shown in Fig.\ \ref{fig:pendantdrop}(a).
We insert two microscale probes into the acoustic cavity to carefully alter the acoustic field within the (horizontal) levitation plane and interact with levitated particles \cite{Brown_VanSaders_Kronenfeld_DeSimone_Jaeger_2024_RSI, Brown_VanSaders_Kronenfeld_DeSimone_Jaeger_2024_GM}.
A levitating droplet of particles is pinned in place by a probe with a large sphere (180 $\upmu$m diam.) that reaches from above into the levitation plane (orange in Fig.\ \ref{fig:pendantdrop}(a)). 
We then move a larger, T-shaped probe within the levitation plane towards the droplet (black in Fig.\ \ref{fig:pendantdrop}(a)). 
The large size of the probe bodies compared to that of the particles implies that the probes generate a force field that attracts the small particles via scattered sound \cite{Brown_VanSaders_Kronenfeld_DeSimone_Jaeger_2024_RSI}.
As the T-shaped probe approaches laterally, the droplet experiences increasing attraction, causing it to deform.
A sample pendant drop is shown in Fig.\ \ref{fig:pendantdrop}(b).
Eventually, the droplet detaches and horizontally ``falls'' onto the T-shaped probe (see Movie S2).
Tracking the center of mass motion of the droplet during this ``falling'' stage, we can determine the effective acceleration $g$ needed for calculating $\mathrm{Bo}$ in Eq.\ \ref{eq:Bondgrav}.     
Figure \ref{fig:pendantdrop}(c) shows a representative plot of droplet velocity vs.\ time after the droplet has detached at $t = 0$.  
Given that the velocity magnitude increases linearly (except very close to impact at the T-shaped probe), the force field generated between the two microprobes to good approximation provides a  constant acceleration $g$ across the drop.
Typical values for $g$ in our experiments are 5-15 m/s$^2$.
Applying this to the early stage of droplet deformation before detachment,  we then fit the drop profile and extract the Bond number.
An example is shown in Fig.\ \ref{fig:pendantdrop}(b) with a best-fit profile overlaid in red, corresponding to a $\mathrm{Bo} = 0.24$.

Figure \ref{fig:pendantdrop}(d) shows interfacial tension data from pendant drop experiments performed at different acoustic pressure values.
Note that these $p_\text{ac}$ are smaller than those cited for free floating droplets; the presence of the microprobes increases the local energy density near the droplet, while the $p_\text{ac}$ values are derived from an empty acoustic cavity.
As $p_{\mathrm{ac}}$ and, thus,  the effective temperature increase, the interfacial tension determined from drop shape also increases, in contrast to typical liquids. 
We can rationalize  this increase of line tension with effective temperature by recalling that  attractive scattering forces and repulsive microstreaming forces both scale in the same way with sound energy density $E_0$ or, equivalently, sound pressure $p_{\mathrm{ac}}$ (Eq. \ref{eq:E0}).  This  means that the particle-particle interaction becomes stiffer while the particle separation remains fixed with increasing  $p_{\mathrm{ac}}$. 
As a result, despite the increased activity inside  droplets  at larger  $p_{\mathrm{ac}}$, their shape reflects a larger interfacial tension.

The scale of the line tension for our acoustically levitated liquid droplets is of the order of $\sim 10$ nN.
To compare to the surface tension of 3D liquids, we obtain an estimate by dividing the $\gamma$ by the particle diameter $d$. This gives a value of $\sim 0.25$ mN/m. 
This is about two orders of magnitude lower than most ordinary liquids, but is comparable to liquid helium  with $\approx 0.3$ mN/m.

\textit{Viscosity.} 
After the droplet ``falls'', it lands on the surface of the larger, T-shaped probe.
It then spreads out and retracts in an oscillatory fashion before settling into a final shape.
When it comes to rest, the sessile droplet appears hydrophilic due to its attraction to the probe. 
Images from the impact of the droplet in Fig.\ \ref{fig:pendantdrop}(b) are shown in Fig.\ S3, and Movie S2 shows an additional example.
We can use this impact to estimate a value for the effective viscosity.
Before impact, the droplet has kinetic and interfacial energy.
As it lands and spreads out, it loses energy via viscous dissipation, and the interfacial energy changes with the drop shape.
We compare the droplet's energy just before impact and at its maximum spread, which allows us to determine a value for the viscosity of a 2D droplet \cite{PasandidehFard_Qiao_Chandra_Mostaghimi_1996, Mao_Kuhn_Tran_1997, Lee_Derome_Guyer_Carmeliet_2016}.
Details of this calculation can be found in the SI.
Note that the units of dynamic viscosity $\eta$ are different in 2D than in 3D, but the units of kinematic viscosity $\nu = \eta/\rho$ are the same.
For this levitated granular droplet, we estimate $\nu \sim 1$ mm$^2$/s, which is very similar to the kinematic viscosity of water as well as to that of a 2D dusty plasma \cite{Nosenko_Goree_2004}.

\section{Liquid pinch-off}
Tracking the breakup of liquid bridges under tension  provides telltale signatures of both the material properties of the liquid and the forces controlling its state \cite{Eggers_Villermaux_2008}. 
Therefore, this type of experiment  offers another opportunity to assess to the effects of the emergent nonreciprocity that generates the liquid-like behavior in the droplets.

\begin{figure}[h]
\centering
\includegraphics[width=\linewidth]{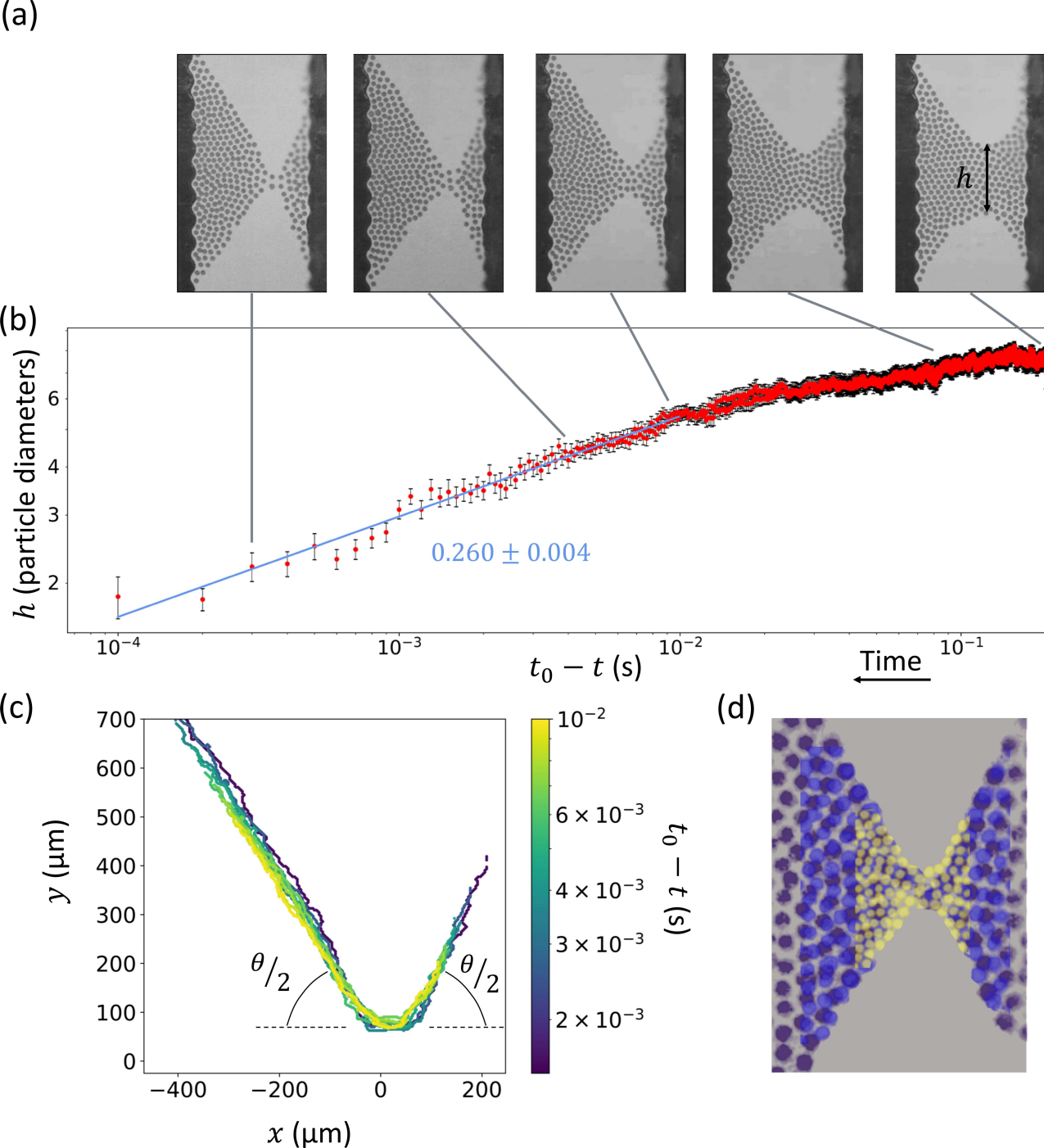}    
\caption{
Active granular liquid bridge pinch-off.
(a) Time series of snapshots from high-speed video.
(b) Ensemble average of neck width $h$ vs.\ time for all experiments; gray lines link the images to the corresponding times in the ensemble plot.
Light blue line is a power law fit to the data on the left side of the plot.
(c) Profiles of the neck, $x$ and $y$ scaled by power law relationship, at times within $10^{-2}$ s of pinch-off (colored by time).
Cone half-angles $\theta/2$ are drawn.
(d) Overlaid images of neck profiles at three times, rescaled by power law relationship (colored by time).
}
\label{fig:pinchoff}
\end{figure}

In our experiments, levitating droplets were anchored to the faces of two opposing T-shaped microprobes.
The configuration was similar to that in Fig.\ \ref{fig:pendantdrop}(a), except that the probe on the right (orange) was replaced by another T-shaped probe (black). Droplets were then stretched by moving one probe slowly ($< 50~\upmu$m/s) laterally until the liquid bridge pinched off. 
This pinch-off process is tracked by measuring the neck width $h$ of the bridge as a function of time $t$ up to pinch-off at $t_0$.
Images from a representative experiment (see Movie S3) are shown in Fig.\ \ref{fig:pinchoff}(a): the dark bars on the left and right sides of each image are the ends of the T-shaped microprobes.
The particles are attracted to the comparatively large faces of these two probes by forces due to scattered sound; indeed, as the panels in Fig.\ \ref{fig:pinchoff}(a) show, the granular liquid very effectively `wets' those probes. By using a corrugated probe face, the contact angle remains fixed as tension is applied.

An ensemble average of neck width $h$ vs.\ time is plotted in Fig.\ \ref{fig:pinchoff}(b).
We performed 28 pinch-off experiments for various drop sizes and acoustic pressure levels, and averaged the value of $h(t)$ for each experiment; lines are drawn from the snapshots of the experiment in (a) to the corresponding times in (b).
Well before pinch-off, the thinning proceeds very slowly at a rate determined by the pulling speed.
Approximately $0.01$ s before pinch-off, or when the neck is close to $5d$ in width, the breakup speeds up.
A fit to the ensemble-averaged data near pinch-off to a power law relationship of the form
\begin{equation}\label{eq:neckpower}
    h \sim (t_0 - t)^{\alpha}
\end{equation}
gives an exponent of $\alpha = 0.260 \pm 0.004$ (if instead  individual traces are fit separately, the mean scaling exponent is $0.28 \pm 0.06$, see Fig.\ S4).

Figure \ref{fig:pinchoff}(c) shows profiles of the upper half of the neck in the 0.01 s before pinch-off (same experiment pictured in (a)).
The shape of the neck shows a double-cone mode with cone angle $\theta \approx 120$-$150^\circ$.
Both the $x$ and $y$ dimensions of the profiles are are scaled by a power law like Eq.\ \ref{eq:neckpower} with $\alpha = 0.26$.
The collapse of these profiles confirms that this stage of breakup is a self-similar process, and that the extent of the neck in the $x$-direction scales similarly to the neck width with time.
Figure \ref{fig:pinchoff}(d) overlays images of the neck at three times, with colors corresponding approximately to the colorbar in (c).
These too mostly collapse onto one profile shape, with some slight variation in the center between the yellow profile (further from pinch-off) and the blue and purple (closer to pinch-off).

Self-similar power law scaling of the neck width near pinch-off is expected for ordinary liquids, where the  exponent $\alpha$ reflects whether viscous, inertial, interfacial, or thermal effects dominate the pinch-off process \cite{Eggers_Villermaux_2008}.
This has been well-studied for 3D systems, with exponents ranging from 1 for breakup dominated by viscous forces to 2/3 for the inviscid case to 0.418 for thermally-driven breakup of nanojets  \cite{Eggers_Villermaux_2008} to 1/3 for breakup via chemical diffusion \cite{Lo_Liu_Mak_Xu_Chao_Li_Shum_Xu_2019}.
2D liquid bridges, however, are fundamentally different.
In particular, while interfacial tension in 3D acts to reduce the cylindrical radius at the narrowest point of the bridge,  in 2D the effect of interfacial tension is to straighten out the neck boundary and thus \emph{increase} its width \cite{Burton_Taborek_2007}.
Though interfacial tension can drive breakup in 2D, it occurs as a secondary effect resulting from the interfacial-tension-driven rebound of drop edges from an extended form \cite{Burton_Taborek_2007}, which gives a scaling exponent $\alpha \approx $ 0.7516.
In our experiments, this rebound cannot occur because the granular liquid bridges remain anchored to the two microprobes. 
Inertia and viscosity are typically not enough to drive breakup in an effectively 2D liquid sheet, instead requiring a contribution from van der Waals (vdW) disjoining pressure or thermal gradients \cite{Wee_Kumar_Liu_Basaran_2024}.
When vdW pressure is considered, it is predicted that a free-standing thin film can collapse with a scaling exponent as low as 2/7.
However, in our experimental system, there is no true analogue for vdW pressure between the interfaces of the neck.
As a result, the data in Fig.\ \ref{fig:pinchoff}(b) cannot be explained by existing scaling laws for pinch-off in 3D or 2D.

\emph{Fluctuation-driven pinch-off.}
The case where pinch-off behavior of a 3D liquid is dominated by the effects of thermal fluctuations seems to be the most similar case to the behavior we observe here.
2D liquid bridge breakup by thermal fluctuations, however, has so far not been explored systematically, and a corresponding exponent $\alpha$ is not known.
Nevertheless, the characteristic double-cone shape observed in 3D liquid threads near pinch-off driven by strong thermal fluctuations \cite{Moseler_Landman_2000, Eggers_2002_thermalfluctuations} suggests that the double-cone-type pinch-off behavior we find (Fig.\ \ref{fig:pinchoff}(a)) is also driven by fluctuations, although in this case the fluctuations arise as a consequence of emergent nonreciprocity.

\begin{figure}[h]
\centering
\includegraphics[width=\linewidth]{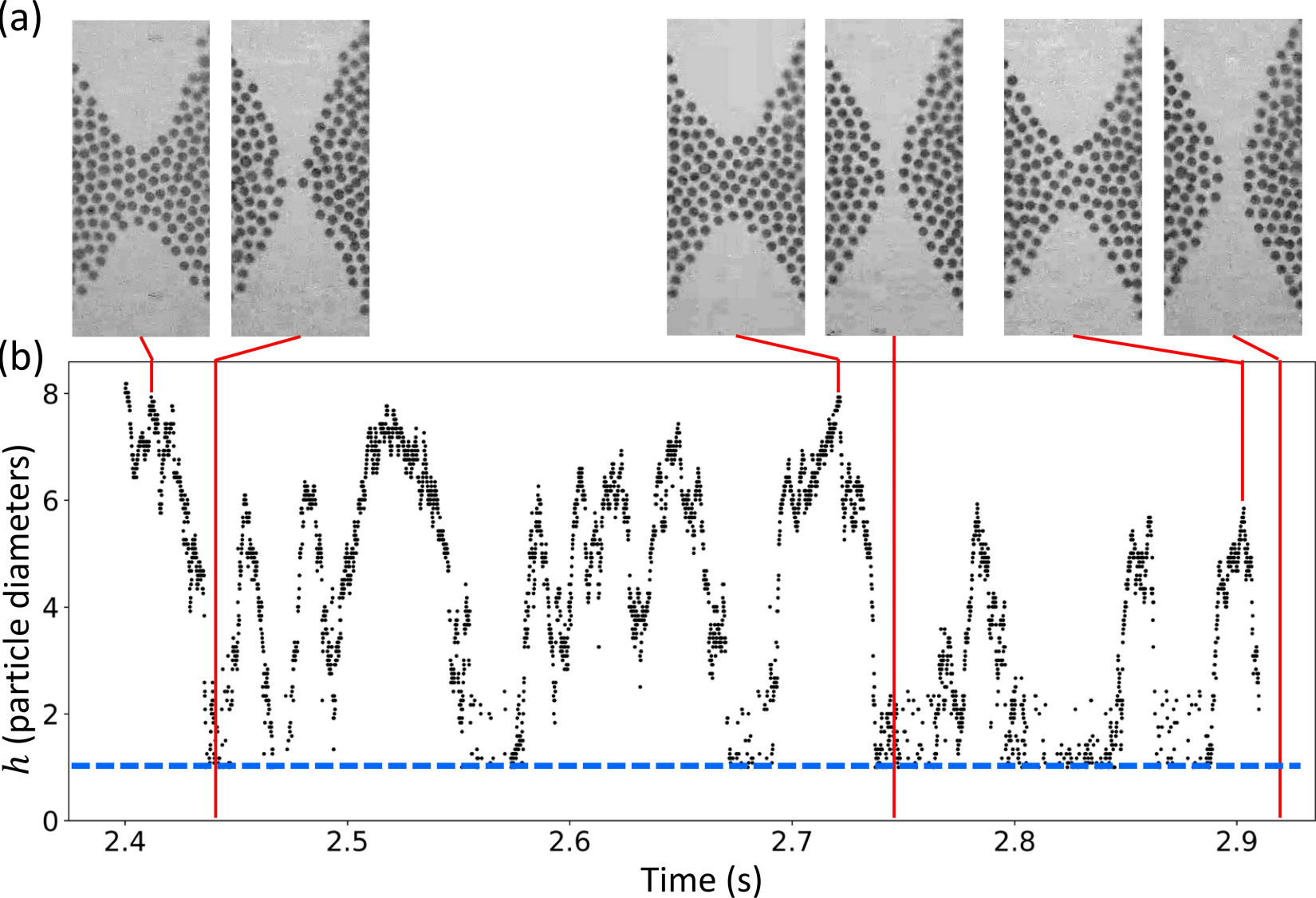}
\caption{
An experimental granular droplet can pinch off and reconnect several times before it pinches off permanently.
(a) A time series of images from the experiment.
(b) Plot of neck width vs.\ time for the end of the experiment where pinch-off and reconnection occurs. Red lines connect images to times. Blue dashed line represents minimum neck width (one particle).
}
\label{fig:reconnect}
\end{figure}

\emph{Pinch-off and re-coalescence.} An unusual feature of this system is the potential for repeated pinch-off and coalescence.
Depending on the experimental parameters, a bridge that pinches off due to particle fluctuations can reconnect, as shown in Fig.\ \ref{fig:reconnect} and Movie S3.
During pinch-off experiments, the microprobe pulling the droplet apart moves very slowly relative to the particle motion.
When the neck is reduced to a width of several particles, fluctuations of individual particles cause the neck to break.
Particles retract to the edges of the remaining, smaller drops.
However, the edges of the two droplet halves may still be just a few particle diameters apart.
If particles from each half experience outward fluctuations, they can get close enough to reconnect the bridge.
The individual particle fluctuations that allow for this reconnection are enabled by the low interfacial tension of the material.
Additionally, there is some acoustic scattering attraction between the two drop halves, encouraging fluctuations towards the midpoint.
Other particles may then follow to again expand the neck to a width of several particles.

This breakup and coalescence cycle can repeat multiple times, lasting until the two droplet halves are too far apart for their respective particles to gain proximity via fluctuations.
Several of these cycles are shown in Fig.\ \ref{fig:reconnect}(a), and a plot of the neck width vs.\ time is shown in (b).
This process exemplifies how the final pinch-off of the granular liquid bridges in this system is driven by particle fluctuations.

\section{Conclusion}
Our results on acoustically levitated granular particles demonstrate that nonreciprocal, multi-body interactions produce a liquid with unique traits.
Increasing the interparticle force strength provokes an increase in particle kinetic energy, as the particle fluctuations are a consequence of the field-mediated interactions.
Asymmetry within the spatial configuration of the particle ensemble gives rise to forces that facilitate particle motion; particles then move into another asymmetric configuration, which incites more motion, and so on.
This phenomenon, enabled by nonreciprocal and multi-body interactions, allows particles to move continuously and form a structure that resembles a liquid droplet.
However, the behavior in this system is athermal, displaying long velocity distribution tails, superdiffusion, and collective motion, all of which are signatures of active and nonreciprocal systems.
A pendant drop of such a liquid deforms similarly to a typical liquid under external stress, but its interfacial tension is low and increases with effective temperature; particles may be moving faster, but the bulk droplet is still bound more tightly.
When this active liquid pinches off, the scaling of the neck width does not align with models of ordinary liquids, and active, athermal fluctuations seem to dominate the behavior.
These particle dynamics also enable a droplet to pinch off and reconnect spontaneously and cyclically.
When particle activity is driven by strongly coupled interactions like these, enhanced activity does not drown out the strengthened effects of those complex interactions, but instead leads to phenomena unique to active liquids with nonreciprocal and multi-body interactions.

\section{Materials and Methods}
\subsection{Experiments}
The experimental setup is pictured in Fig.\ \ref{fig:acousticlevitation}(a).
We drive a piezoelectric ultrasound transducer (Steiner \& Martins SMBLTD45F40H) with a 40.45-40.7 kHz sinusoidal signal from a function generator (BK Precision 4052) and a high-voltage amplifier (A.A. Lab Systems A-301 HS).
The transducer is attached to a custom-machined aluminum horn, which amplifies the signal (horn design from \cite{Andrade_Buiochi_Adamowski_2010}).
The horn is heated to a constant temperature of 35$^\circ$ C to maintain consistent resonance properties.
The bottom edge of the horn is approximately one half-wavelength $\sim 4.2$ mm from a transparent reflector plate.
The reflector is made of transparent glass and coated with ITO to minimize the influence of charging and electrostatic forces.
A mirror is placed below the reflector at a 45$^\circ$ angle, and a Phantom v2512 high speed camera is used to capture experiments from below and observe the nodal plane of the cavity.
A Navitar Resolv4K lens is used to achieve a maximum resolution of approximately 2.5 $\upmu$m per pixel at a working distance of $6.5$ cm.
The acoustic pressure in the system is monitored by several low-profile ultrasound sensors (PUI Audio SMUT-1040K-TT) placed below the reflector.
A feedback control system with the ultrasound sensors and the function generator maintains a constant acoustic pressure in the cavity during experiments, as described in \cite{Brown_VanSaders_Kronenfeld_DeSimone_Jaeger_2024_RSI}, which helps to accommodate small changes in the lab environment.
The setup is partially enclosed in an acrylic box and located atop a passive vibration isolation platform (ThorLabs IsoPlate PTT900600).

Polystyrene spheres with diameter $d = 40.4 \pm 0.9$ $\upmu$m were levitated in these experiments (microParticles GmbH).
These particles are coated with fumed silica nanoparticles (Evonik Aeroxide AluC) to prevent contact adhesion from van der Waals forces; this is described in \cite{Wu_Esposito_Mao_Jaeger_2025}.
To avoid the influence of electrostatic forces, levitated particles and microprobes were discharged prior to each experiment with a soft x-ray photoionizer (Hamamatsu L12645).
Microprobes  with a width of approximately 200 $\upmu$m are used to perturb the system; these were manufactured using a continuous liquid interface production method \cite{Hsiao_Lee_Samuelsen_Lipkowitz_Kronenfeld_Ilyn_Shih_Dulay_Tate_Shaqfeh_et_al._2022}.
A motorized micromanipulator (Eppendorf Patchman NP2) is used to control microprobe motion.

\subsection{Analysis}
All image processing and data analysis was performed in Python. Raft and microprobe positions were obtained from experimental images using the Segment Anything Model \cite{Kirillov_Mintun_Ravi_Mao_Rolland_Gustafson_Xiao_Whitehead_Berg_Lo_et_al._2023}.
Particle positions and tracks were determined using the trackpy Python package \cite{CROCKER1996298}.
Droplet center of mass motion was removed from the particle positions so as to examine the the relative motion of particles within the droplet.
Fitting of pendant drop profiles utilized a custom modified version of the OpenDrop software \cite{Berry_Neeson_Dagastine_Chan_Tabor_2015, Huang_Skoufis_Denning_Qi_Dagastine_Tabor_Berry_2021}.

The local orientational order parameter $\psi_6$ is defined for a single particle $j$ as
\begin{equation}
   \psi_{6, j} = \frac{1}{n} \sum_\ell e^{6 i \theta_\ell}
\end{equation}
where the sum is over a particle's $n$ nearest neighbors and $\theta_\ell$ is the angle of the vector between particles $j$ and $\ell$. 
This is a complex vector quantity, where the magnitude suggests degree of crystallinity and the angle suggests crystal lattice orientation.
To study trends of drop disorder with effective temperature, we compute $|\psi_6| = \langle |\psi_{6, j}|\rangle_j$ for particles in the interior of each droplet for each frame, then time-average $|\psi_6|$ over the course of the experiment.

The static structure factor $S(\mb{k})$ can be computed directly from particle positions as
\begin{equation}
\begin{aligned}
    S(\mb{k})
    =
    \frac{1}{N} \Bigg \langle \sum_{i = 1}^N \sum_{j = 1}^N &\big [ \cos \p{\mb{k} \cdot \mb{r}_i} \cos \p{\mb{k} \cdot \mb{r}_j} 
    \\
    &+ \sin \p{\mb{k} \cdot \mb{r}_i} \sin \p{\mb{k} \cdot \mb{r}_j}\big ] \Bigg \rangle
\end{aligned}
\end{equation}
where $\mb{r}_i$ is the position of particle $i$ \cite{Thorneywork_Schnyder_Aarts_Horbach_Roth_Dullens_2018}.
This calculation is performed by creating a 2D grid with spacing $2\pi/L$ ($L$: droplet diameter) and computing $S(\mb{k})$ for all $\mb{k} = (k_x, k_y)$ for each pair of particles.
These values are averaged for each $k = |\mb{k}|$ to obtain $S(k)$.

\section{Acknowledgements}
We thank Qinghao Mao, Brady Wu, Jochem Meijer, Alice Pelosse, and Billie Meadowcroft for inspiring discussions. 
We thank Jason Kronenfeld and Joe DeSimone for manufacturing the microprobe tips.
This work was supported by the NSF through DMR-2104733. The
experiments utilized the shared experimental facilities at the
University of Chicago MRSEC, which is funded by the
National Science Foundation under award number DMR-2011854. The research also benefited from computational
resources and services supported by the Research Computing Center at the University of Chicago.

\bibliographystyle{abbrv}
\bibliography{draft}

\end{document}